\documentclass[aps,pre,twocolumn,groupedaddress,superscriptaddress,showpacs]{revtex4-1}
\usepackage{epsfig}
\usepackage{amsmath}

\begin{document}

\title{Scaling Relations for Watersheds}

\author{E. Fehr}
\email{ericfehr@ethz.ch}
\author{D. Kadau}
\author{N. A. M. Ara{\'u}jo}
\affiliation{IfB, ETH Z\"urich, 8093 Z\"urich, Switzerland}

\author{J. S. Andrade Jr.}
\author{H. J. Herrmann}
\affiliation{IfB, ETH Z\"urich, 8093 Z\"urich, Switzerland}
\affiliation{Departamento de F\'{\i}sica, Universidade Federal do Cear\'a, 60451-970 Fortaleza, Cear\'a, Brazil}

\begin{abstract}
We study the morphology of watersheds in two and three dimensional systems subjected to different degrees of spatial correlations. The response of these objects to small, local perturbations is also investigated with extensive numerical simulations. We find the fractal dimension of the watersheds to generally decrease with the Hurst exponent, which quantifies the degree of spatial correlations. Moreover, in two dimensions, our results match the range of fractal dimensions $1.10 \leq d_f \leq 1.15$ observed for natural landscapes. We report that the watershed is strongly affected by local perturbations. For perturbed two and three dimensional systems, we observe a power-law scaling behavior for the distribution of areas (volumes) enclosed by the original and the displaced watershed, and for the distribution of distances between outlets. Finite-size effects are analyzed and the resulting scaling exponents are shown to depend significantly on the Hurst exponent. The intrinsic relation between watershed and invasion percolation, as well as relations between exponents conjectured in previous studies with two dimensional systems, are now confirmed by our results in three dimensions.
\end{abstract}


\maketitle 

\section{Introduction}

Watersheds are the lines separating adjacent drainage basins (catchments). They play a fundamental role in water management~\cite{Vorosmarty:1998aa,Kwarteng:2000aa,Sarangi:2005aa}, landslides \cite{Dhakal:2004aa,Pradhan:2006zr,Lazzari:2006aa,Lee:2006aa}, and flood prevention \cite{Lee:2006aa,Burlando:1994ys,Yang:2007aa}. Natural watersheds are fractal \cite{Breyer:1992sp}. Geographers and geomorphologists have found the evolution of watersheds to be driven by local events classified as stream captures or drainage rearrangements.These events can affect the biogeography \cite{Burridge:2007fk}, and may occur due to various mechanisms like erosion \cite{Garcia-Castellanos:2009fk,Linkeviciene:2009uq,Bishop:1995zr}, natural damming \cite{Lee:2006aa}, tectonic motion \cite{Garcia-Castellanos:2003uq,Dorsey:2006kx,Lock:2006vn}, as well as volcanic activity \cite{Beranek:2006ys}.

Fractality of watersheds was first claimed in Ref. \cite{Breyer:1992sp}. Their observations were limited to small scales and only few samples. In Ref. \cite{Fehr:2009aa}, extensive numerical simulations were devised to study watersheds on uncorrelated artificial landscapes, as well as on large-scale natural landscapes in the form of digital elevation maps (DEM), as the ones obtained from satellite imagery \cite{Farr:2007aa}. Defining the watershed to be the line dividing the entire landscape into two parts, a novel and efficient identification algorithm was developed. Using this method, the self-similarity was confirmed and the fractal dimension was estimated to be $1.10\pm0.01$ for the Alps, $1.11\pm0.01$ for the Himalayas, and $1.211\pm0.001$ for uncorrelated artificial landscapes. Due to the ubiquity of the obtained fractal dimension, also relations to other physical models have been proposed, such as optimal paths and optimal path cracks \cite{Andrade:2009vn,Andrade:2011aa,Oliveira:2011fk}, bridge percolation \cite{Cieplak:1994kx,Cieplak:1996aa,Araujo:2011xx}, and the surface of explosive percolation clusters \cite{Araujo:2010ys,Schrenk:2011fk}. This opens a broad range of possible implications and applications of the properties of watersheds. In this paper, we study the effect of correlations on the fractal dimension of watersheds for artificial landscapes and compare in two dimensions to the values obtained for several natural landscapes ranging from flat (Big Lakes) to mountainous regions (Rocky Mountains).

The mechanisms triggering the stream capture and drainage rearrangement events \cite{Garcia-Castellanos:2009fk,Linkeviciene:2009uq,Bishop:1995zr,Lee:2006aa,Garcia-Castellanos:2003uq,Dorsey:2006kx,Lock:2006vn,Beranek:2006ys}, although diverse, are all modifications of the topography. In that perspective, Fehr {\it et al.} \cite{Fehr:2011aa} investigated the effect of such events by applying small local perturbations to natural and artificial landscapes and found that watersheds can be strongly affected. Power-law scaling behavior was observed, e.g. for the distribution of the area enclosed by the original and the displaced watershed, and shown to be independent on the strength of the perturbation. Tuning the correlation degree in the artificial landscapes, the values of the exponents were matched with the ones obtained for natural landscapes. Additionally, for uncorrelated artificial landscapes, relations for the scaling laws to properties of invasion percolation \cite{Araujo:2005qe} were conjectured. We now extend these studies to three dimensional systems, where the watershed is a surface dividing the volume into two parts. We focus mainly on the study of conjectured scaling relations, analyze the finite-size effects, and clarify the relation to invasion percolation.

This paper is organized as follows. In Section~\ref{SECfrac} we briefly revisit the definition of our model and show the dependence of the fractal dimension on the landscape correlations for both two and three dimensions. Section~\ref{SEC2Dpert} summarizes our results for the power-law distributions found for the effects of topographical modifications \cite{Fehr:2011aa}. To validate the relations conjectured for two dimensions, we study in Section~\ref{SEC3Dpert} the effect of perturbations for three dimensional systems. Conclusions are drawn in Section~\ref{SECconclusion}.

\begin{figure*}[!ht] 
\centering
\begin{tabular}{cc}
{\bf\Large (a)} & {\bf\Large (b)}\\
\epsfig{figure=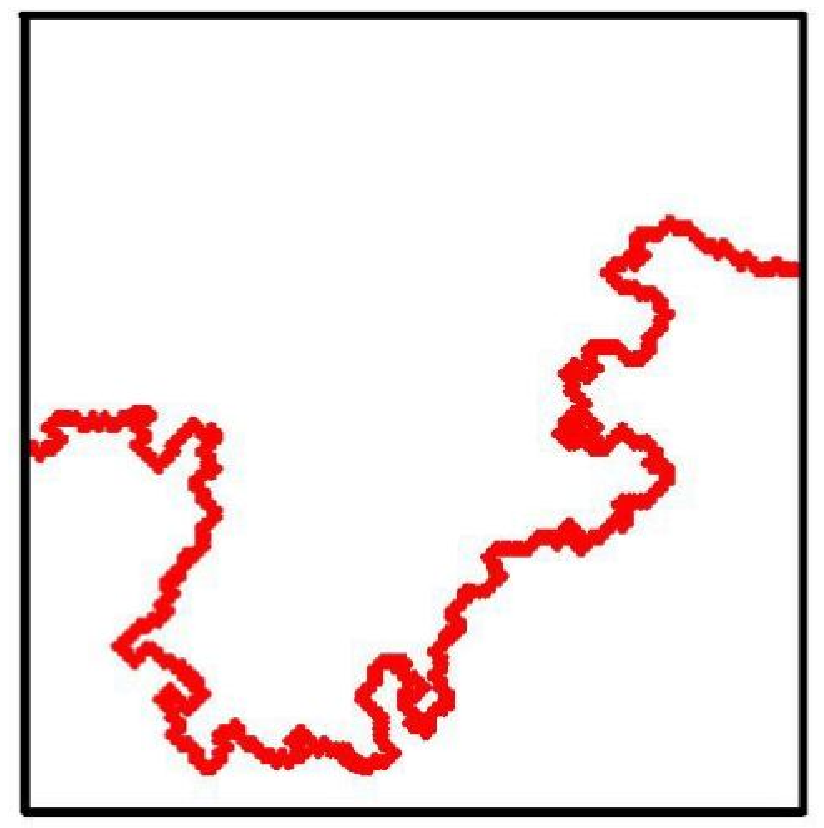,width=8.3cm} &\epsfig{figure=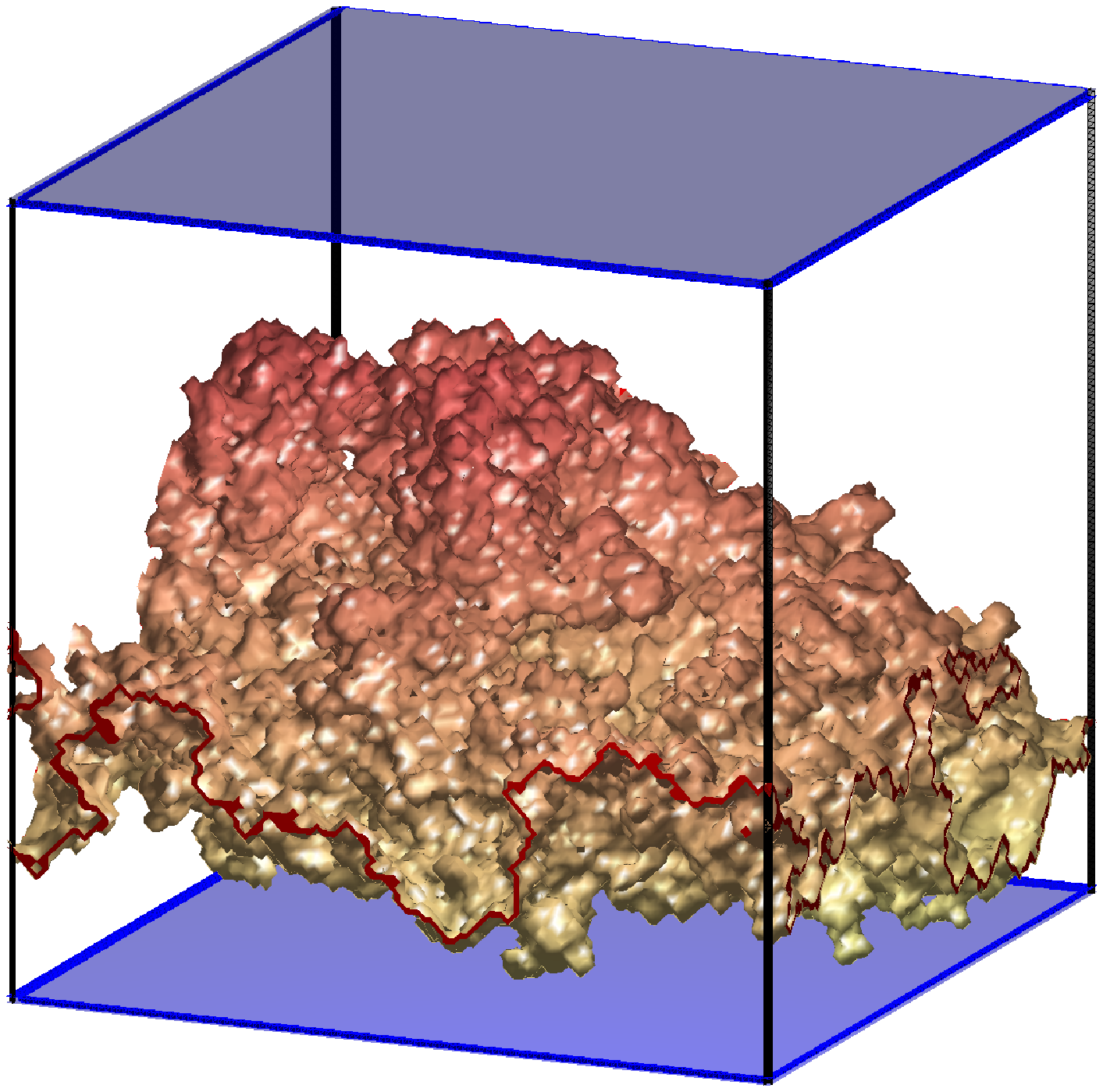,width=8.3cm}
\end{tabular}
\caption{(color online) Example of a watershed for uncorrelated systems with linear size $L=129$ lattice sites in {\bf(a)} two and {\bf(b)} three dimensions. The upper and lower catchments drain to the top and bottom boundary, respectively.}\label{fig1}
\end{figure*}

\section{The Watershed and its Fractal Structure}\label{SECfrac}

In nature, watersheds are the lines separating adjacent drainage basins (catchments). We study the watershed lines in two dimensions as well as their extension to three dimensions. We use real and artificial systems in the form of Digital Elevation Maps (DEM), giving the elevations (resistances) on a square (cubic) lattice. We define as the watershed, the line (surface) dividing the entire system into two sets of sites, \emph{catchments}, as shown in Fig.~\ref{fig1}. Each site in a catchment drains to the same boundary of a chosen pair of opposite boundaries (top-bottom in Fig.~\ref{fig1}) of the DEM. Hereby, the drainage of a site is determined by following the local slope and filling the valleys (local minima), until eventually reaching one of the two boundaries. For the determination of this line we use an iterative invasion percolation procedure (IIP), as introduced in Ref. \cite{Fehr:2009aa}.

Recently, watersheds have numerically been shown to be self-similar objects \cite{Fehr:2009aa}. These structures are typically characterized by their fractal dimension $d_f$, which is defined through the scaling of the mass $M$, corresponding to the number of sites or bonds in the watershed, with the linear system size $L$,
\begin{equation}\label{EQfractaldim}
M\sim L^{d_f}\,\, .
\end{equation}
In Ref. \cite{Fehr:2009aa} the fractal dimension of watersheds for uncorrelated artificial landscapes has been estimated to be $d_f=1.211\pm0.001$, which differs from the ones obtained for landscapes from satellite imagery shown in Tab.~\ref{Tab_Df}. This is expected as in natural landscapes typically long-range correlations are present \cite{Pastor-Satorras:1998ys}. Therefore, we study here the dependence of the fractal dimension on correlations for several artificial landscapes.

\begin{table}[h!]
\caption{Fractal dimension of watersheds for natural landscapes obtained from satellite imagery \cite{Farr:2007aa}. We added here the values presented in Ref. \cite{Fehr:2009aa} for the Alps and the Himalayas for completeness. The fractal dimensions are obtained using the yardstick method. The error bars are of the order of $2\%$.}\label{Tab_Df}
\centering
\begin{tabular}{l|l}
\hline
\hline
Landscape & $d_f$\\
\hline
Alpes & 1.10 \cite{Fehr:2009aa}\\
Europe & 1.10\\
Rocky Mountains & 1.11\\
Himalayas & 1.11 \cite{Fehr:2009aa}\\
Kongo & 1.11\\
Andes & 1.12\\
Appalachians & 1.12\\
Brazil & 1.12\\
Germany & 1.14\\
Big Lakes & 1.15\\
\hline
\hline
\end{tabular}
\end{table}

Spatial long-range correlated distributions can be obtained with fractional Brownian motion (fBm) \cite{Mandelbrot:1967zr, Peitgen:1988aa}. Similarly to previous studies \cite{Fehr:2011aa,Morais:2011eu,Sahimi:1994uq,Sahimi:1996kx,Makse:1996vn,Oliveira:2011fk, Prakash:1992ly,Kikkinides:1999ve,Stanley:1999qf,Makse:2000bh,Araujo:2002dq,Araujo:2003cr,Du:1996nx}, we use the Fourier filtering method \cite{Peitgen:1988aa,Sahimi:1994uq,Sahimi:1996kx,Makse:1996vn,Oliveira:2011fk}, which allows to control the nature and the strength of correlations. A detailed description of this method can be found, e.g., in Ref. \cite{Oliveira:2011fk}. In brief, the desired correlated distribution can be introduced by generating Fourier coefficients in the reciprocal space according to a power-law spectral density. 
For each frequency in the reciprocal space, we calculate these Fourier coefficients through a random phase in the interval $\left.\left[0:2\pi\right.\right)$ and an amplitude $\left(\sqrt{k_1^2+\dots+k_d^2}\right)^{-2H-d}$, where $k_i$ are the frequency indices of the discrete Fourier transform, $d$ the spatial dimension, and $H$ the Hurst exponent. The inverse Fourier transform is applied to obtain the distribution in real space. Finally, we normalize the spatial domain distribution in the range $[0:1]$ to represent the correlated topology, characterized by the Hurst exponent $H$. Three different categories of fBm surfaces can be distinguished: $0<H<1/2$, $H=1/2$, and $1/2<H<1$ \cite{Sahimi:1995fk}. The correlations between the increments are persistent (positive) for $1/2 < H < 1$, meaning that sites with similar height tend to cluster together, leading to rather smooth surfaces. The opposite is true for $0 < H < 1/2$, where the correlations of the increments are anti-persistent (negative), resulting in surfaces that seem to oscillate more erratically. For $H = 1/2$, the classical Brownian motion is recovered, where the increments are uncorrelated but the obtained heights are still correlated. The uncorrelated distribution of heights is solely obtained for a constant spectral density, i.e. $H = -d/2$ ($H = -1$ and $H=-3/2$ in two and three dimensions, respectively).

\begin{figure}[!ht] 
\centerline{\epsfig{figure=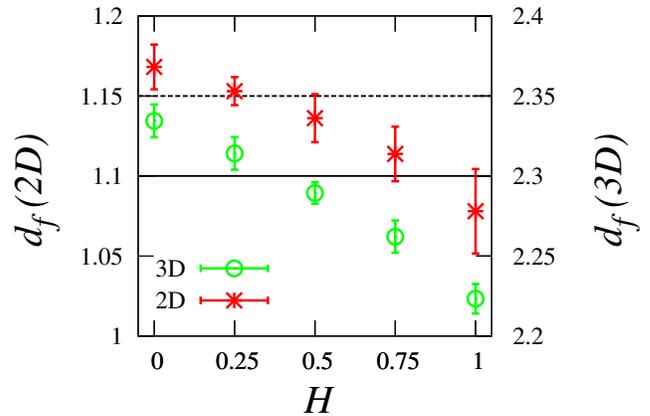,width=8.3cm}}
\caption{(color online) The fractal dimension $d_f$ of the watershed as a function of the Hurst exponent $H$ of the system in two (red stars) and three dimensions (green open circles) according to left and right hand axes, respectively. In two dimensions, each point corresponds to an estimate of the fractal dimension by fitting the power law in Eq.~(\ref{EQfractaldim}) to the watershed masses for system sizes $L=\{5,\, 17,\, 65,\, 257,\, 1025,\, 4097\}$. For each system size, the mass was averaged over $10^4$ landscape realizations. In three dimensions, the system sizes $L=\{5,\, 9,\, 17,\, 33,\, 65,\, 129,\, 257\}$ were used with the same number of realizations. The lines show the fractal dimension of the watershed obtained for the Alps (solid) and close to the Big Lakes (dotted), characterizing, according to the left hand axes, the range of estimates for natural landscapes as summarized in Tab.~\ref{Tab_Df}. In the typical range of natural landscapes, $0.3<H<0.5$, the simulation results are in agreement with the natural ones.}\label{fig2}
\end{figure}

In Fig.~\ref{fig2} we plot the dependency of the fractal dimension of the watershed on the Hurst exponent, measuring the degree of correlations. A monotonic decrease of $d_f$ with $H$ is observed, in line with what was observed for the optimal path crack in $d=2$ \cite{Oliveira:2011fk}. Considering the typical range of $0.3 < H < 0.5$ for natural landscapes \cite{Pastor-Satorras:1998ys}, our simulation results in two dimensions are in agreement with the values for the fractal dimension of watersheds in natural landscapes measured from satellite imagery, as listed in Tab.~\ref{Tab_Df}. The lines in Fig.~\ref{fig2} show the fractal dimension obtained for the watershed of the Alps (solid line) and the one close to the Big Lakes (dotted line), characterizing the range of values obtained for natural landscapes.

We now extend the concept of a watershed to a three dimensional system, in which the values at the sites no longer represent heights but for instance resistances. The watershed is now a surface, as shown in Fig.~\ref{fig1}b, that divides the system into two parts. Similarly, each site in one part drains to the same boundary of a chosen pair of opposite boundaries (top-bottom in Fig.~\ref{fig1}) of the DEM. The drainage of a site is determined by following the lowest gradient in the resistances and filling the regions of local minima, until eventually reaching one of the two boundaries. Again, the IIP procedure \cite{Fehr:2009aa} can be used to determine the watershed surface numerically.
Alike the two dimensional case, we use Eq.~(\ref{EQfractaldim}) to estimate the fractal dimension. For uncorrelated three dimensional systems, we find $d_f=2.48\pm0.02$. In Fig.~\ref{fig2}, we show that the fractal dimension of the watershed for three dimensional systems also decreases continuously with the Hurst exponent $H$.

\section{Impact of Perturbations in Two Dimensions}\label{SEC2Dpert}

So far we have considered that the properties of the landscape, and consequently of the watershed, are static, i.e., do not change in time. However, landscapes might change due to several phenomena such as erosion, tectonic motion, and volcanic activity. Such changes in the landscape are known to trigger local events called stream capture \cite{Garcia-Castellanos:2003uq,Garcia-Castellanos:2009fk,Linkeviciene:2009uq,Dorsey:2006kx,Lock:2006vn,Beranek:2006ys,Bishop:1995zr}, which can affect the watershed. In what follows, we extend the recent work presented in Ref. \cite{Fehr:2011aa}.
%
\begin{figure*}[!ht] 
\centering
\begin{tabular}{cc}
{\bf\Large (a)} & {\bf\Large (b)}\\
\epsfig{figure=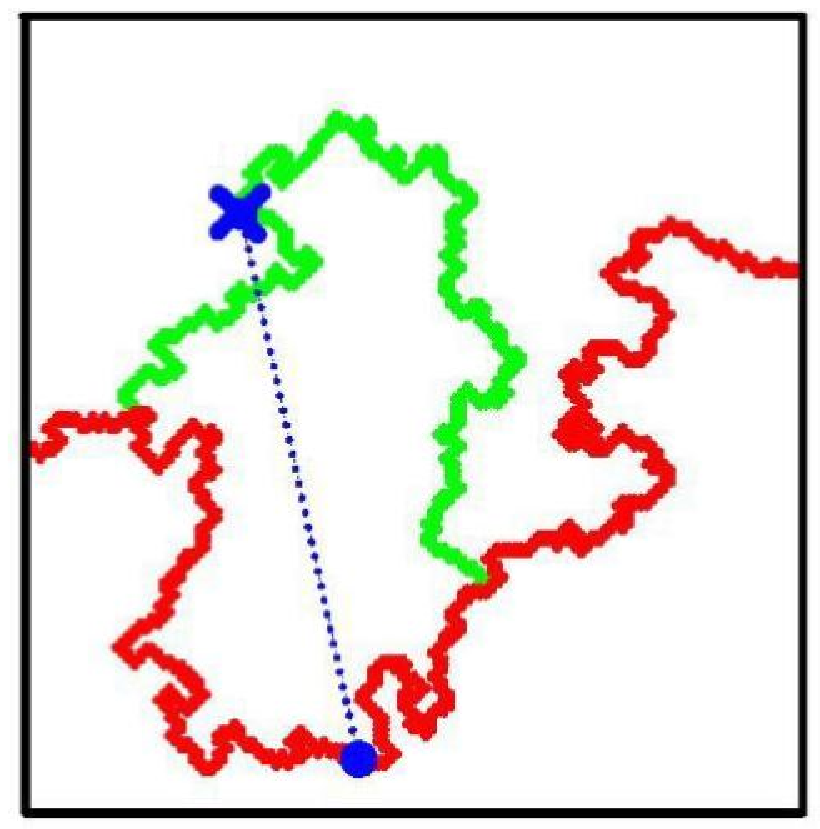,width=8.3cm} & \epsfig{figure=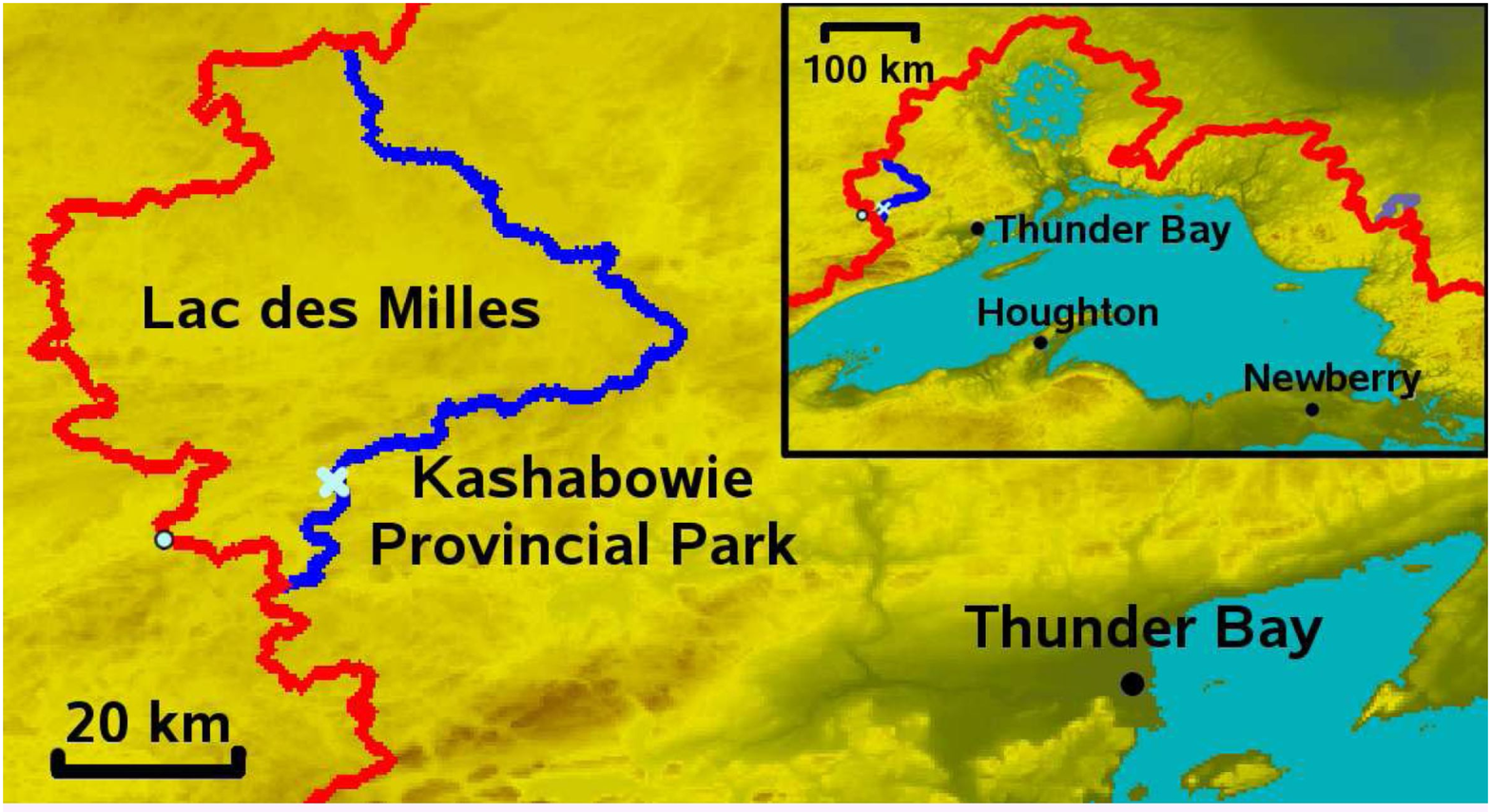,width=9.7cm}
\end{tabular}
\caption{(color online) {\bf(a)} Example of a displacement (top, green) of the original watershed (bottom, red, the same as in Fig.~\ref{fig1}a) after a perturbation at the spot marked by a cross for an uncorrelated artificial landscape. The dotted line connects the two \emph{outlets} of the area before (cross) and after (dot) perturbation and defines the distance $R$. {\bf(b)} The watershed close to the big lakes in the United States (light (red) line). A perturbation of 2 m at a spot (cross) near Thunder Bay caused a change in the watershed (dark (blue) line). The watershed displacement encloses an area of about 3730 km$^2$ \cite{Fehr:2011aa}. The dot marks the new \emph{outlet} of the area after the perturbation.}\label{figEGpert}
\end{figure*}
%
This approach is based on a perturbation scheme, where a local event is induced by changing the height $h_k$ at a single site $k$ of the system to $h_k+\Delta$, with $\Delta$ being the perturbation strength. It has been shown in Ref. \cite{Fehr:2011aa} that, without loss of generality, we can fix the perturbation strength to the height difference between the highest and lowest height of the landscape. As shown in Fig.~\ref{figEGpert}b, we quantify this response by the area, i.e. the number of sites, $N_s$, enclosed by the resulting watershed (dark (blue) line) after the perturbation at site $k$ (cross) and the watershed of the original landscape (light (red) line). The water can only escape from this area through one single site, which we call \emph{outlet}. In this scheme, two \emph{outlets} can be considered, one before (cross) and another after the perturbation (dot). The former always coincides with the perturbed site $k$ and can be connected to the latter, inside the enclosed area, by an invasion percolation cluster, whose mass we denote by $M$. The number of enclosed sites $N_s$, the mass $M$ of the connecting cluster and the distance $R$ between the two outlets (dotted line in Fig.~\ref{figEGpert}a) are measured. After that, the original landscape is restored by resetting the height at $k$ to its initial value. Except of those sites located on the original watershed, this procedure is repeated for every site $k$ in the landscape. In the following, we consider only perturbations actually leading to a displacement of the watershed, i.e. $N_s>0$. To reduce finite-size effects, we explicitly exclude those perturbations, where the changed areas touch the borders of the system and, therefore, the original and the perturbed watershed are always overlapping at the boundaries. From the obtained set of measures $N_s$, $M$, and $R$, we calculate the distribution $P(N_s)$ of the numbers of enclosed sites (areas) $N_s$, the distribution $P(M)$ of the clusters mass, and the probability distribution $P(R)$ of the Euclidean distance $R$ between the two outlets. To investigate the dependence of $N_s$ and $M$ on the distance $R$, we define the average $\left<N_s\right>$ and distribution $P(N_s|R)$ of areas associated with a distance $R$, as well as the average $\left<M\right>$ and distribution $P(M|R)$. For the dependence of $M$ and $N_s$ on each other we study the average mass associated with an area $N_s$. All these distributions and measures were sampled for each configuration, then averaged over 2000 realizations of systems with size $L=513$ and over 4000 for $L=129$ and $257$. In Ref. \cite{Fehr:2011aa}, the distributions $P(N_s)$, $P(R)$, $P(N_s|R)$, and $P(M|R)$, as well as $\left<N_s\right>$, have been shown to follow power laws of the form
\begin{subequations}
\begin{eqnarray}
P(N_s) &\sim& N_s^{-\beta},\label{EQ_PA}\\
P(R) &\sim& R^{-\rho},\label{EQ_PR}\\
\left<N_s\right> &\sim& R^{\sigma},\label{EQ_AR}\\
P(N_s|R) &\sim& N_s^{-\alpha},\label{EQ_PAR}\\
P(M|R) &\sim& M^{-(1+\alpha^*)},\label{EQ_PMR}
\end{eqnarray}
\end{subequations}
where the exponents $\alpha$, $\beta$, and $\rho$ have been estimated for several natural as well as artificial landscapes. While $\rho$ and $\beta$ were found to increase with increasing $H$, the exponent $\alpha$ did decrease. It has been observed that the average angle between the lines connecting the outlets with the center of mass of the area, decreases with increasing correlation degree. This implies that, on average, the two outlets approach each other with increasing $H$, so that $R$ is no longer representative of the area extension. Therefore, with $H\rightarrow 1$, fixing the distance $R$ no longer restricts the areas entering the distribution $P(N_s|R)$, such that the exponent $\alpha$ decreases with $H$ and approaches $\beta$. For all considered landscapes the value $\sigma=2$ has been obtained. The distribution $P(M|R)$ for uncorrelated landscapes can be directly related to point-to-point invasion percolation, with $\alpha^*=1.39$ being the subcritical exponent as introduced in Ref. \cite{Araujo:2005qe}.

\begin{figure}[!ht] 
\centerline{\epsfig{figure=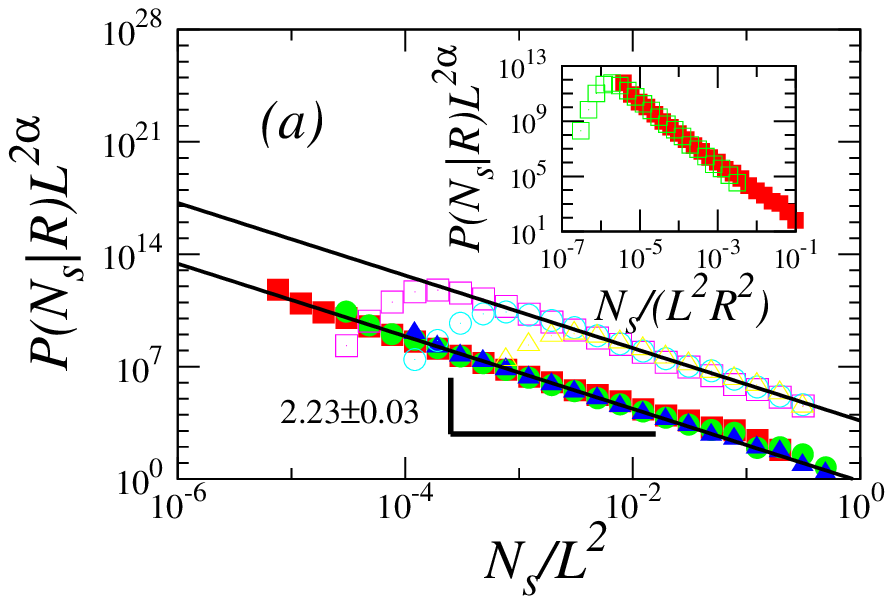,width=8.3cm}}
\centerline{\epsfig{figure=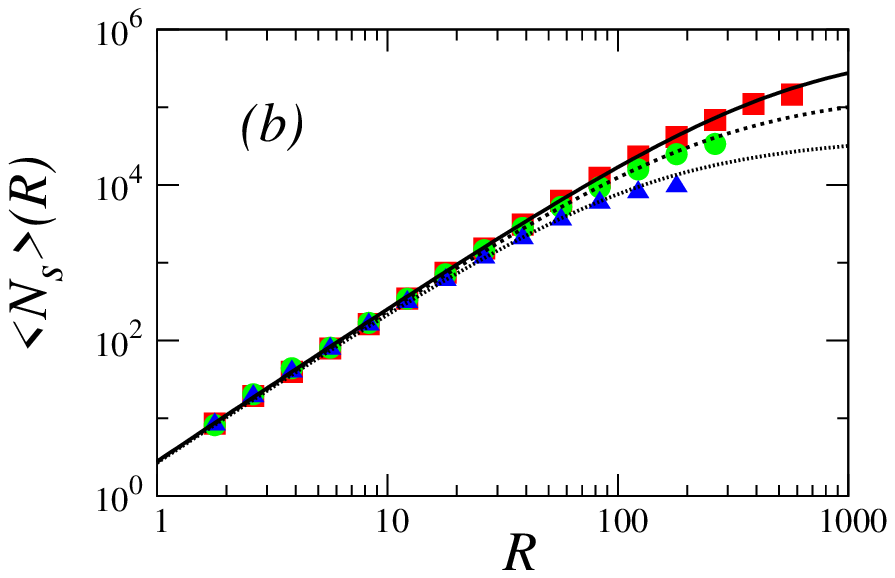,width=8.3cm}}
\caption{(color online) {\bf(a)} Data collapse for the distribution $P(N_s|R)$ with $R=1$ (filled) and $R=10$ (open) and system sizes $L=\{129,\, 257,\, 513\}$ (triangles, circles, and squares, respectively), using the scaling $P(N_s|R)=L^{2\alpha}f[N_sL^2]$. The solid lines represent fits to the data of a power law with exponent $\alpha=2.23\pm0.03$. The inset shows the collapse of the same data for $L=513$, when the x-axis is rescaled by $R^d$, visualizing the lower cutoff from the power-law behavior. {\bf(b)} The average number of sites (area) $\left<N_s\right>$ enclosed by the original and the perturbed watershed at an outlet distance $R$ for the same landscapes and system sizes as in (a). The lines show the expression given by Eq.~(\ref{EqARpredict}). Each data point, in both {\bf(a)} and {\bf(b)}, corresponds to an average over 2000 realizations of linear system size $L=513$ and over 4000 for $L=129$ and $257$. The error bars are smaller than the size of the symbols.}\label{fig2D_PAR}
\end{figure}

In the following we present a functional description for the average number of enclosed sites $\left<N_s\right>(R,L)$ that captures the dependence on $R$, including also the finite-size effects. We can write the average for a given distance $R$ as the first moment of the distribution $P(N_s|R)$, i.e.,
\begin{equation}\label{EQ1stmoment}
\left<N_s\right>(R) = \int\mathrm{d}N_s\, N_s\, P(N_s|R) \,\, .
\end{equation}
The distribution obtained from our simulation has a lower and an upper cutoff, as can be seen from Fig.~\ref{fig2D_PAR}a. Considering that those perturbations with areas touching the boundary are excluded, we need to determine the scaling of the cutoffs numerically. From the data collapse achieved using the scaling $P(N_s|R)=L^{2\alpha}f[N_sL^2]$, we can see that the upper cutoff indeed scales as $L^d$. Similarly, the lower cutoff follows $R^d$ as obtained from the data collapse shown in the inset of Fig.~\ref{fig2D_PAR}a. Applying both cutoffs as bounds for the integral on the right hand side of Eq.~(\ref{EQ1stmoment}), as well as for the normalization integral for $P(N_s|R)$, we obtain
\begin{eqnarray}\label{EqARpredict}
\displaystyle\left<N_s\right>(R,L) &=& \frac{\displaystyle\int_{R^{d}}^{L^{d}}\mathrm{d}N_s\, N_s^{1-\alpha}}{\displaystyle\int_{R^{d}}^{L^{d}}\mathrm{d}N_s\, N_s^{-\alpha}} \\\nonumber
&=& \displaystyle C\left(\frac{1-\alpha}{2-\alpha}\right)\, \left(\frac{L^{d(2-\alpha)}-R^{d(2-\alpha)}}{L^{d(1-\alpha)}-R^{d(1-\alpha)}}\right)\,\, .
\end{eqnarray}
This result matches, within the error bars, our simulation data as shown in Fig.~\ref{fig2D_PAR}b. In the limit \mbox{$L\rightarrow\infty$}, only possible because $\alpha>2$ \cite{Fehr:2011aa}, it reduces to \mbox{$\left<N_s\right>(R)=C(1-\alpha)/(2-\alpha) R^{d}$} and, therefore, \mbox{$\sigma=2$}, as observed numerically.

In a similar way, as calculated for the number of enclosed sites $N_s$, we investigate the mass $M$ of the invasion percolation cluster which connects the two outlets before and after perturbation. The cluster is always a subset of the enclosed sites $N_s$. Hence, we assume also the distribution $P(M)$ and the average mass $\left<M\right>$ to obey power laws of the form
\begin{subequations}
\begin{eqnarray}
P(M) &\sim& M^{-(1+\beta^*)}\label{EQ_PM}\\
\mathrm{and}\qquad \left<M\right> &\sim& R^{\sigma^*},\label{EQ_MR}
\end{eqnarray}
\end{subequations}
where we choose Eq.~(\ref{EQ_PM}) to have a similar form as Eq.~(\ref{EQ_PMR}). Analogously to Eq.~(\ref{EQ1stmoment}), we can define the average mass $\left<M\right>$ as the first moment of the distribution $P(M|R)$. With a (finite-size) scaling analysis similar to the one used to get the cutoffs in $P(N_s|R)$, we find the lower and upper cutoff of $P(M|R)$ to scale as $R^{d_f^*}$ and $L^{d_f^*}$, respectively, where $d_f^*$ is the fractal dimension of the invasion percolation clusters. Using Eq.~(\ref{EQ_PMR}) together with Eq.~(\ref{EQ1stmoment}) with $M$ instead of $N_s$ we obtain
\begin{eqnarray}\label{EqMRpredict}
\left<M\right>(R,L) &=& \frac{\displaystyle\int_{R^{d_f^*}}^{L^{d_f^*}}\mathrm{d}M\, M^{-\alpha^*}}{\displaystyle\int_{R^{d_f^*}}^{L^{d_f^*}}\mathrm{d}M\, M^{-\alpha^*-1}} \\\nonumber
&=& \displaystyle C\left(\frac{\alpha^*}{\alpha^*-1}\right)\, \left(\frac{L^{{d_f^*}(1-\alpha^*)}-R^{{d_f^*}(1-\alpha^*)}}{L^{-{d_f^*}\alpha^*}-R^{-{d_f^*}\alpha^*}}\right)\,\, .
\end{eqnarray}
%
\begin{figure}[!ht] 
\centerline{\epsfig{figure=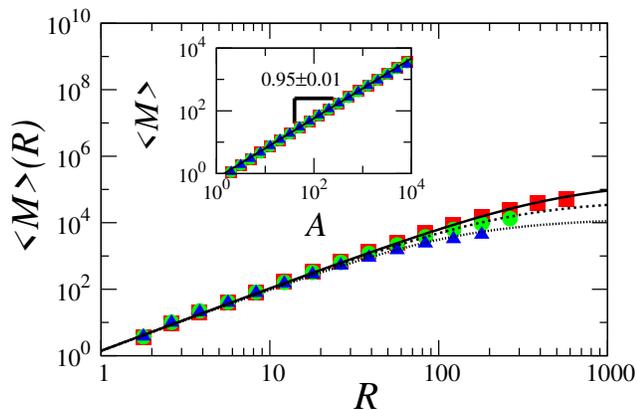,width=8.3cm}}
\caption{(color online) The average mass $\left<M\right>$ of the invasion percolation cluster connecting the two outlets with a distance $R$ for uncorrelated two dimensional landscapes of sizes $L=\{129,\, 257,\, 513\}$ (triangles, circles, and squares, respectively). The lines show the predictions according to Eq.~(\ref{EqMRpredict}) for different system sizes. The inset shows the average mass varying with the number of enclosed sites. The line is a power-law fit to the data yielding an exponent $0.95\pm0.01\approx d_f^*/2$. Each data point corresponds to an average over 2000 realizations of systems with size $L=513$ and over 4000 for $L=129$ and $257$. The error bars are smaller than the size of the symbols.}\label{fig2D_MR}
\end{figure}
%
\begin{figure}[!ht] 
\centerline{\epsfig{figure=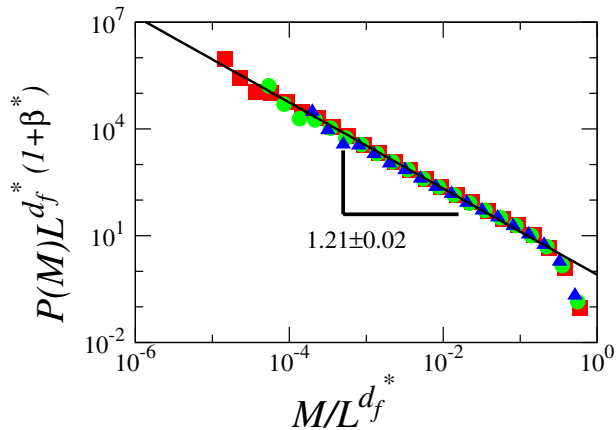,width=8.3cm}}
\caption{(color online) Data collapse of the size distribution $P(M)$ of the mass $M$ of the invasion percolation cluster connecting the two outlets for uncorrelated landscapes of three different system sizes $L=\{129,\, 257,\, 513\}$ (triangles, circles, and squares, respectively). The line represents a power-law fit to the data for the largest landscape (squares) yielding an exponent $1+\beta^*=1.21\pm0.02$. Each data point is an average over 2000 realizations of systems with size $L=513$ and over 4000 for $L=129$ and $257$. The error bars are smaller than the size of the symbols.}\label{fig2D_PM}
\end{figure}
%
In Fig.~\ref{fig2D_MR} the matching, within the error bars, of this result with the simulation data is shown. For $L\rightarrow\infty$ the result reduces to \mbox{$\left<M\right>(R)=\alpha^*/(\alpha^*-1)R^{d_f^*}$}, as $\alpha^* > 1$, yielding $\sigma^*=d_f^*=91/48$. The relation of the area and the mass is analyzed with the average mass $\left<M\right>$ associated to a given number of enclosed sites $N_s$. From the dependence of $\left<N_s\right>$ and $\left<M\right>$ on $R$ we expect
\begin{equation}\label{EQ_MA}
\left<M\right> \sim N_s^{\frac{d_f^*}{\sigma}}\,\, ,
\end{equation}
which matches the value $0.95\pm0.01$ obtained from the power-law fit to the data plotted in the inset of Fig.~\ref{fig2D_MR}. Based on this result, using Eq.~(\ref{EQ_MA}) together with Eq.~(\ref{EQ_PA}) and (\ref{EQ_PM}) we expect \mbox{$\beta^*=\sigma\beta/d_f^*-1\approx0.22$}. This prediction matches the value $\beta^*=0.21\pm0.02$ estimated from the data collapse of $P(M)$ shown in Fig.~\ref{fig2D_PM}. Summarizing our results in two dimensions, together with the relations conjectured in Ref. \cite{Fehr:2011aa} we obtain the relations
\begin{subequations}
\begin{eqnarray}
\alpha&=&\frac{d_f^*}{\sigma}(1+\alpha^*), \label{EQ_alphaRel}\\
\beta&=&\frac{d_f^*}{\sigma}(1+\beta^*), \label{EQ_betaRel}\\
\rho&=&\sigma\beta, \label{EQ_rhoRel}
\end{eqnarray}
\end{subequations}
between the exponents $\alpha$ and $\beta$ for the numbers of sites $N_s$ enclosed by the original and the perturbed watershed, $\alpha^*$ and $\beta^*$ for the mass $M$ of the invasion percolation clusters, and $\rho$ for the probability density to induce, after perturbation, a change at a given distance. Despite the similarity of $\alpha$ and $\rho$ for uncorrelated two dimensional landscapes \cite{Fehr:2011aa}, they differ significantly in three dimensions, as shown in the next section.

\section{Impact of Perturbations in three Dimensions}\label{SEC3Dpert}

\begin{figure*}[!ht]
\centering
\begin{tabular}{cc}
{\bf\Large (a)} & {\bf\Large (b)}\\
\epsfig{figure=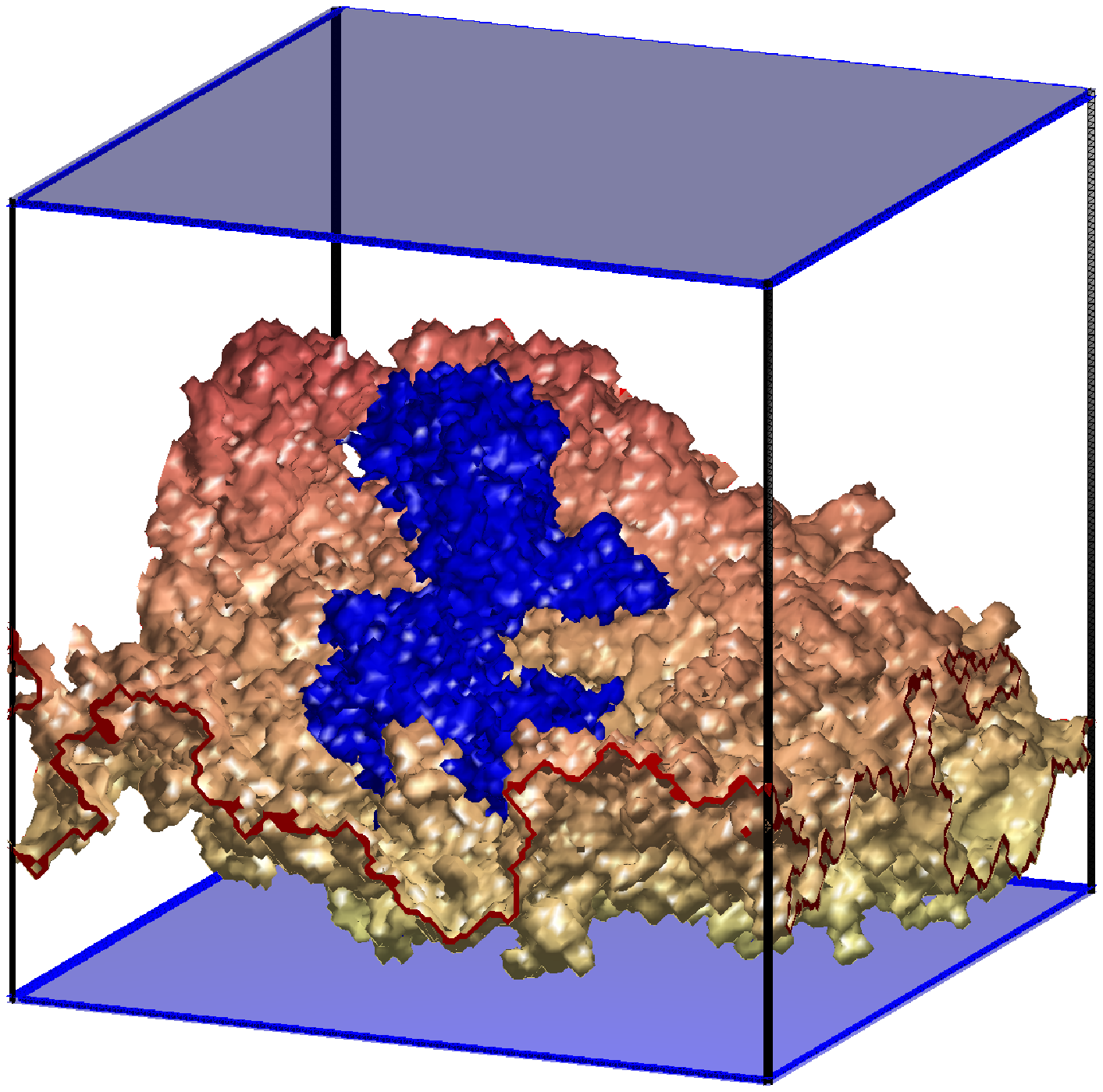,width=8.3cm} & \epsfig{figure=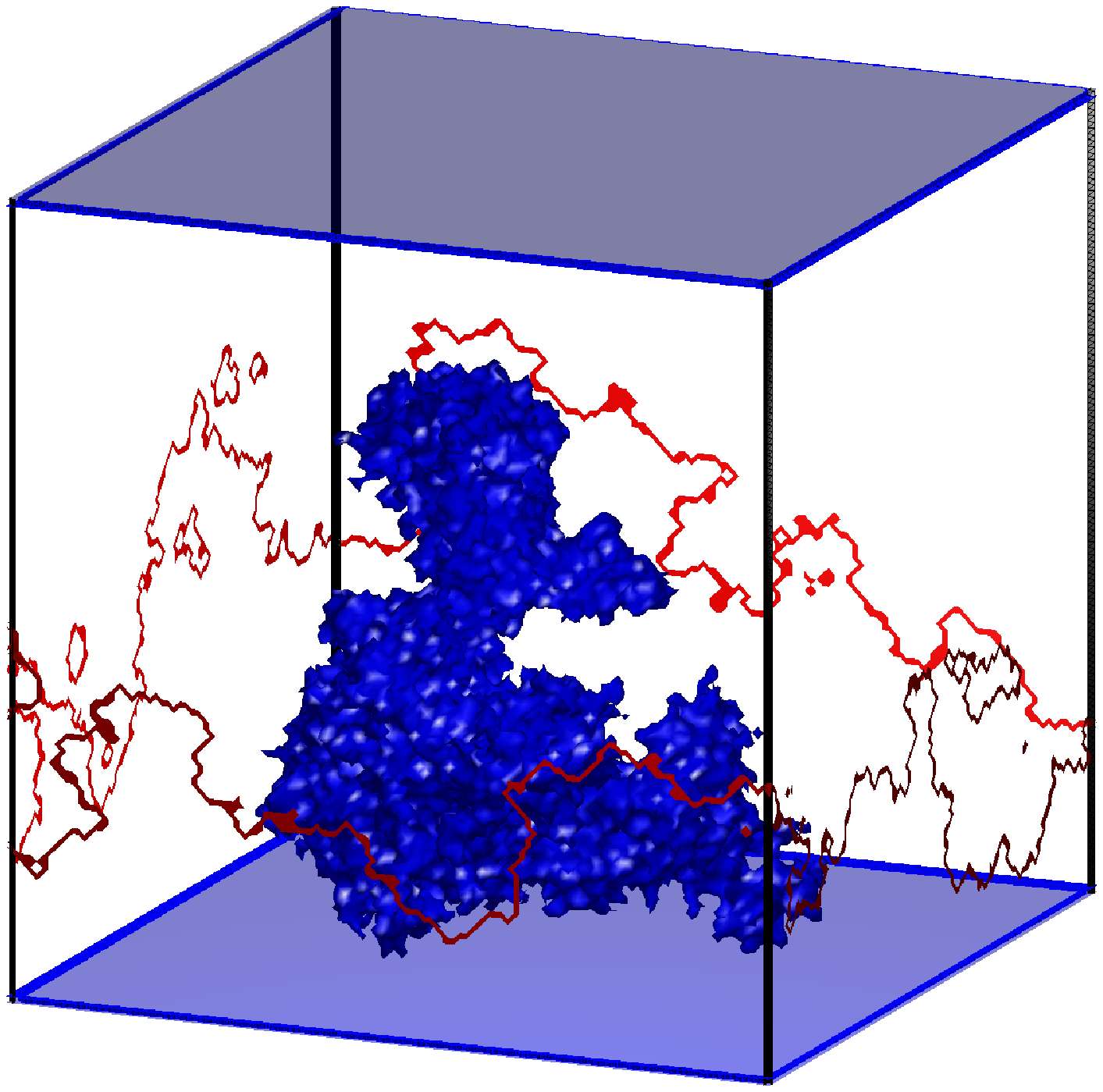,width=8.3cm}
\end{tabular}
\caption{(color online) {\bf (a)} The largest of all observed changed volumes (dark, blue) attached to the perturbed watershed (light, red) for the same uncorrelated three dimensional system of linear size $L=129$, as used in Fig.~\ref{fig1}b. The upper and lower catchments drain to the top and bottom borders, respectively. {\bf (b)} The same changed volume as in (a) without the watershed for better visibility. The lines mark the sites where the original watershed intersects the system boundaries.}\label{figSingleExmp}
\end{figure*}

\begin{figure}[!ht] 
\centerline{\epsfig{figure=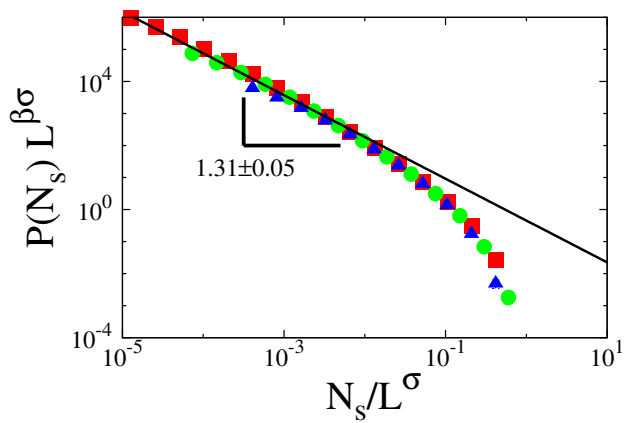,width=8.3cm}}
\caption{(color online) Data collapse of the distribution $P(N_s)$ of the number of sites (volume) enclosed by the original and the perturbed watershed for uncorrelated three dimensional systems of size $L=\{33,\, 65,\, 129\}$, (squares, circles, triangles, respectively). The line represents a power-law fit to the data yielding an exponent $\beta=1.31\pm0.05$. Each data point is an average over $2000$ realizations. The error bars are smaller than the size of the symbols.}\label{fig3D_PA}
\end{figure}

In this section we extend the previous concepts to three dimensional systems. As introduced in Sec.~\ref{SECfrac}, the watershed in three dimensions is a surface that divides the system into two parts. Similar to two dimensions, a perturbation is induced by changing the local resistance $r_k$ at site $k$ to $r_k+\Delta$, where $\Delta$ is the perturbation strength. We quantify the impact on the watershed surface, as before, by the number of sites $N_s$ enclosed by the original and the perturbed watershed, which now corresponds to a change in volume. As an example, the largest of all changed volumes observed in our simulations is shown in Fig.~\ref{figSingleExmp}. The outlets of this volume, before and after perturbation, are determined as well as their distance $R$ and mass $M$ of the invasion percolation cluster connecting them. The original system is then restored by resetting $r_k$ to its initial value. This procedure is repeated for all sites $k$ of the system, except those located on the watershed. Again, the averages $\left<N_s\right>$, $\left<M\right>$, and distributions $P(R)$, $P(N_s)$, $P(N_s|R)$, $P(M)$, and $P(M|R)$ are sampled and averaged over $2000$ configurations. Similar to the two dimensional case, as shown in Fig.~\ref{fig3D_PA} for the distribution $P(N_s)$ of volumes, we find these quantities to follow power laws of the form introduced in Eqs.~(\ref{EQ_PA})-(\ref{EQ_PMR}) and Eqs.~(\ref{EQ_PM})-(\ref{EQ_MR}). The values of all the exponents estimated for uncorrelated systems are summarized in Tab.~\ref{TAB_exp}.
%
%
%
\begin{table}[h!]
\caption{Summary of the exponents numerically obtained in this study. Estimates are given for watersheds on natural and uncorrelated artificial landscapes in two dimensions, as well as for uncorrelated artificial systems in three dimensions. For exponents obtained in previous works the corresponding citations are given. The range $1.10-1.15$ of fractal dimension for natural landscapes is obtained from Tab.~\ref{Tab_Df}.}\label{TAB_exp}
\centering
\begin{tabular}{c|c|r@{$\pm$}ll|r@{$\pm$}ll|r@{$\pm$}ll}
\hline
\hline
Type& Eq. &\multicolumn{3}{|c|}{natural}&\multicolumn{6}{|c}{artificial (uncorr.)}\\
\hline
$d$&&\multicolumn{2}{|c}{2}& &\multicolumn{2}{|c}{2}&&\multicolumn{2}{|c}{3}&\\
$d_f$& (\ref{EQfractaldim}) &\multicolumn{2}{|c}{1.10 - 1.15}& &1.211&0.001&\cite{Fehr:2009aa}&2.48&0.02&\\
$\alpha$&(\ref{EQ_PAR})&2.3&0.2&\cite{Fehr:2011aa}&2.23&0.03&\cite{Fehr:2011aa}&2.4&0.1&\\
$\beta$&(\ref{EQ_PA})&1.65&0.15&\cite{Fehr:2011aa}&1.16&0.03&\cite{Fehr:2011aa}&1.31&0.05&\\
$\rho$&(\ref{EQ_PR})&3.1&0.3&\cite{Fehr:2011aa}&2.21&0.01&\cite{Fehr:2011aa}&3.2&0.2&\\
$\sigma$&(\ref{EQ_AR})&\multicolumn{2}{|c}{2}&\cite{Fehr:2011aa}&\multicolumn{2}{|c}{2}&\cite{Fehr:2011aa}&2.45&0.05&\\
$\alpha^*$&(\ref{EQ_PMR})&\multicolumn{2}{|c}{--}&&1.39&0.03&\cite{Araujo:2005qe,Fehr:2011aa}&1.4&0.1&\\
$\beta^*$&(\ref{EQ_PM})&\multicolumn{2}{|c}{--}&&0.21&0.02&&0.29&0.06&\\
$\sigma^*$&(\ref{EQ_MR})&\multicolumn{2}{|c}{--}&&\multicolumn{2}{|c}{$91/48$}&\cite{Stauffer:1994aa}&\multicolumn{2}{|c}{2.53 \cite{Stauffer:1994aa}}&\\
\hline
\hline
\end{tabular}
\end{table}

\begin{figure}[!ht] 
\centerline{\epsfig{figure=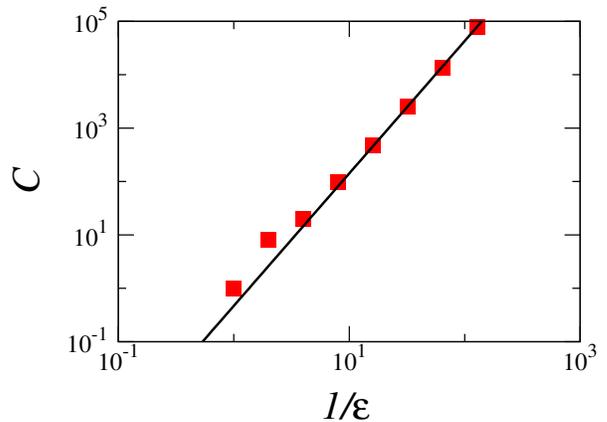,width=8.3cm}}
\caption{(color online) Number of cubic boxes $C$ of size $\varepsilon$ covering the changed volume, shown in Fig.~\ref{figSingleExmp}, as obtained from a box counting method. Each data point consists of a single measurement. The line corresponds to $\varepsilon^{2.48}$.}\label{figBoxSingle}
\end{figure}

Although the volumes $N_s$ are still compact, we find $\sigma=2.48\pm0.02$ by an analysis of $\left<N_s\right>$ and $P(N_s|R)$ similar to the one performed in two dimensions using Eq.~(\ref{EqARpredict}) in the limit $L\rightarrow\infty$. The obtained value of $\sigma$ is close to the fractal dimension of the watershed, what suggests that the mass of the changed volume is dominated by its surface, in contrast to what we observe in two dimensions. This is also confirmed by a box counting analysis of the largest of all changed volumes, as depicted in Fig.~\ref{figBoxSingle}.
%
%
In agreement with the findings by Lee in Ref. \cite{Lee:2009fk}, that the size distribution of sub-critical point-to-point invasion percolation, as introduced by Ara{\'u}jo {\it et al.} \cite{Araujo:2005qe}, is independent on the dimensionality of the system, we obtain $\alpha^*=1.4\pm0.1$ for uncorrelated systems, similar to the value found in two dimensions. This confirms the relation to invasion percolation drawn in Ref. \cite{Fehr:2011aa}. Inserting the estimates of $d_f^*,\,\sigma$, and $\alpha^*$, given in Tab.~\ref{TAB_exp}, into Eq.~(\ref{EQ_alphaRel}), we find $\alpha\approx2.48$ matching with the value $2.4\pm0.1$ obtained numerically, hence validating the conjectured relation. Similarly, we also observe that our results are consistent with Eq.~(\ref{EQ_betaRel}).

\begin{figure}[!hc] 
\centerline{\epsfig{figure=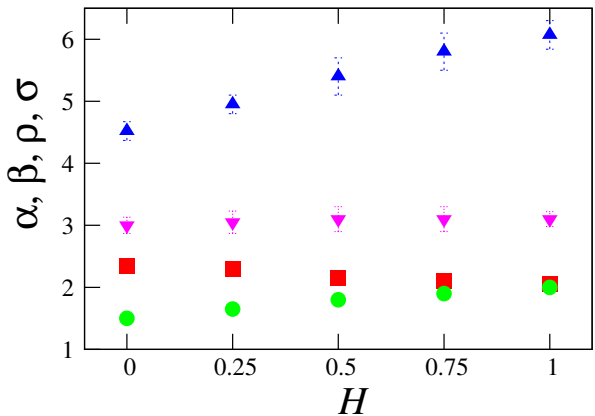,width=8.3cm}}
\caption{(color online) The exponents $\alpha,\,\beta,\,\rho,\,\sigma$ (squares, circles, triangles up, and triangles down, respectively) as a function of the Hurst exponent $H$ for perturbations in 3D. Each data point consists of a similar measurement as performed in Fig.~\ref{fig3D_PA} to obtain $\beta$ for uncorrelated systems.}\label{fig3D_exp}
\end{figure}

The dependence of the exponents $\alpha$, $\beta$, and $\rho$ on the Hurst exponent $H$ is shown in Fig.~\ref{fig3D_exp}, confirming the relation $\rho=\sigma\beta$ (Eq.~(\ref{EQ_rhoRel})), independent on the degree of correlations. We observe $\alpha$ to decrease for $H>0$ and to approach $\beta$, similar as found in two dimensions (compare Section~\ref{SEC2Dpert}).
%
%

\section{Conclusion}\label{SECconclusion}

In summary, we were able to show that watersheds are fractals with a fractal dimension in the range of $1.10\leq d_f \leq 1.15$ for all analyzed natural landscapes from mountainous (e.g. Alps) to rather flat ones (e.g. Big Lakes). By studying model landscapes with long-range correlations characterized by the Hurst exponent $H$, we determined the dependence of the fractal dimension on $H$ in both two and three dimensions, where for the former we found good quantitative agreement with natural landscapes, for which $0.3 < H < 0.5$. Extending the work done in Ref. \cite{Fehr:2011aa} on the impact of perturbations on watersheds in two dimensional systems, we found the lower and upper bounds of the distribution of areas to scale with distance and system size with an exponent $\sigma=2$ in two and $\sigma=2.48\pm0.02$ in three dimensions. Considering these cutoffs, we were able to derive a function describing the average area, which includes the finite-size effects. In a similar way, by integrating $P(M|R)$, we obtained a function for the average mass $\left<M\right>$ of the invasion percolation cluster.
In the extension of the perturbation study to three dimensional systems, we observed the changed volumes to be dominated by their surface, i.e., the watershed, resulting in a value of $\sigma$ close to the fractal dimension of the watershed itself. The distributions and averages for the three dimensional case are shown to follow power laws as in two dimensions. Indeed, we found in the uncorrelated case an intrinsic relation to invasion percolation. Finally, our results in three dimensions are consistent with the conjectured relations between the exponents. As a followup of the work presented here, it would be interesting to see how the impact of perturbations could be related to other physical models, such as optimal path cracks \cite{Andrade:2009vn,Oliveira:2011fk}, bridge percolation \cite{Araujo:2011xx}, and the surface of explosive percolation clusters \cite{Araujo:2010ys,Schrenk:2011fk}.

We acknowledge useful discussions with D. Garcia-Castellanos, C. Moukarzel, L. Hurni, and J. Schrenk and thank CNPq, CAPES, FUNCAP, and the CNPq/FUNCAP-Pronex grant for financial support.


\begin{thebibliography}{99}
\bibitem{Vorosmarty:1998aa} C. J. Vorosmarty, C. A. Federer, and A. L. Schloss, J. Hydrol. 207, 147 (1998).
%
\bibitem{Kwarteng:2000aa} A. Kwarteng, M. Viswanathan, M. Al-Senafy, and T. Rashid, J. Arid. Environ. 46, 137 (2000).
%
\bibitem{Sarangi:2005aa} A. Sarangi and A. K. Bhattacharya, Agric. Water Man- age. 78, 195 (2005).
%
\bibitem{Dhakal:2004aa} A. S. Dhakal and R. C. Sidle, Hydrol. Process. 18, 757 (2004).
%
\bibitem{Pradhan:2006zr} B. Pradhan, R. P. Singh, and M. F. Buchroithner, Adv. Space Res. 37, 698 (2006).
%
\bibitem{Lazzari:2006aa} M. Lazzari, E. Geraldi, V. Lapenna, and A. Loperte, Landslides 3, 275 (2006).
%
\bibitem{Lee:2006aa} K. T. Lee and Y.-T. Lin, J. Am. Water Resour. Assoc. 42, 1615 (2006).
%
\bibitem{Burlando:1994ys} P. Burlando, M. Mancini, and R. Rosso, IFIP Transac- tions B (Applications in Technology) 16, 91 (1994).
%
\bibitem{Yang:2007aa} D. Yang, Y. Zhao, R. Armstrong, D. Robinson, and M.- J. Brodzik, J. Geophys. Res. 112, F02S22 (2007).
%
\bibitem{Breyer:1992sp} S. P. Breyer and R. S. Snow, Geomorphology 5, 143 (1992).
%
\bibitem{Burridge:2007fk} C. P. Burridge, D. Craw, and J. M. Waters, Mol. Ecol. 16, 1883 (2007).
%
\bibitem{Garcia-Castellanos:2009fk} D. Garcia-Castellanos, F. Estrada, I. Jimenez-Munt, C. Gorini, M. Fernandez, J. Verges, and R. De Vicente, Nature (London) 462, 778 (2009).
%
\bibitem{Linkeviciene:2009uq} R. Linkeviciene, Holocene 19, 1233 (2009).
%
\bibitem{Bishop:1995zr} P. Bishop, Prog. Phys. Geogr. 19, 449 (1995).
%
\bibitem{Garcia-Castellanos:2003uq} D. Garcia-Castellanos, J. Verges, J. Gaspar-Escribano, and S. Cloetingh, J. Geophys. Res. 108, 2347 (2003).
%
\bibitem{Dorsey:2006kx} R. J. Dorsey and J. J. Roering, Geomorphology 73, 16 (2006).
%
\bibitem{Lock:2006vn} J. Lock, H. Kelsey, K. Furlong, and A. Woolace, Geol. Soc. Am. Bull. 118, 1232 (2006).
%
\bibitem{Beranek:2006ys} L. P. Beranek, P. K. Link, and C. M. Fanning, Geol. Soc. Am. Bull. 118, 1027 (2006).
%
\bibitem{Fehr:2009aa} E. Fehr, J. S. Andrade, Jr., S. D. da Cunha, L. R. da Silva, H. J. Herrmann, D. Kadau, C. F. Moukarzel, and E. A. Oliveira, J. Stat. Mech.-Theory Exp. , P09007 (2009).
%
\bibitem{Farr:2007aa} T. G. Farr, P. A. Rosen, E. Caro, R. Crippen, R. Duren, S. Hensley, M. Kobrick, M. Paller, E. Rodriguez, L. Roth, D. Seal, S. Shaffer, J. Shimada, J. Umland, M. Werner, M. Oskin, D. Burbank, and D. Alsdorf, Rev. Geophys. 45, 33 (2007).
%
\bibitem{Andrade:2009vn} J. S. Andrade, Jr., E. A. Oliveira, A. A. Moreira, and H. J. Herrmann, Phys. Rev. Lett. 103, 225503 (2009).
%
\bibitem{Andrade:2011aa} J. S. Andrade, Jr., S. D. S. Reis, E. A. Oliveira, E. Fehr, and H. J. Herrmann, Comput. Sci. Eng. 13, 74 (2011).
%
\bibitem{Oliveira:2011fk} E. A. Oliveira, K. J. Schrenk, N. A. M. Ara{\'u}jo, H. J. Herrmann, and J. S. Andrade, Jr., Phys. Rev. E 83, 046113 (2011).
%
\bibitem{Cieplak:1994kx} M. Cieplak, A. Maritan, and J. R. Banavar, Phys. Rev. Lett. 72, 2320 (1994).
%
\bibitem{Cieplak:1996aa} M. Cieplak, A. Maritan, and J. R. Banavar, Phys. Rev. Lett. 76, 3754 (1996).
%
\bibitem{Araujo:2011xx} N. A. M. Ara{\'u}jo, K. J. Schrenk, J. S. Andrade, Jr., and H. J. Herrmann, arXiv:1103.3256v1.
%
\bibitem{Araujo:2010ys} N. A. M. Ara{\'u}jo and H. J. Herrmann, Phys. Rev. Lett. 105, 035701 (2010).
%
\bibitem{Schrenk:2011fk} K. J. Schrenk, N. A. M. Ara{\'u}jo, and H. J. Herrmann, arXiv:1104.5376.
%
\bibitem{Fehr:2011aa} E. Fehr, D. Kadau, J. S. Andrade, Jr., and H. J. Herrmann, Phys. Rev. Lett. 106, 048501 (2011).
%
\bibitem{Araujo:2005qe} A. D. Ara{\'u}jo, T. F. Vasconcelos, A. A. Moreira, L. S. Lucena, and J. S. Andrade, Jr., Phys. Rev. E 72, 041404 (2005).
%
\bibitem{Pastor-Satorras:1998ys} R. Pastor-Satorras and D. H. Rothman, Phys. Rev. Lett. 80, 4349 (1998).
%
\bibitem{Mandelbrot:1967zr} B. B. Mandelbrot, Science 156, 636 (1967).
%
\bibitem{Peitgen:1988aa} H. Peitgen and D. Saupe, eds., {\it The Science of Fractal Images} (Springer, New York, 1988).
%
\bibitem{Morais:2011eu} P. A. Morais, E. A.Oliveira, N. A. M. Ara{\'u}jo, H. J. Herrmann, and J. S. Andrade, Jr., arXiv:1102.5615.
%
\bibitem{Sahimi:1994uq} M. Sahimi, J. Phys. I France 4, 1263 (1994).
%
\bibitem{Sahimi:1996kx} M. Sahimi and S. Mukhopadhyay, Phys. Rev. E 54, 3870 (1996).
%
\bibitem{Makse:1996vn} H. A. Makse, S. Havlin, M. Schwartz, and H. E. Stanley, Phys. Rev. E 53, 5445 (1996).
%
\bibitem{Prakash:1992ly} S. Prakash, S. Havlin, M. Schwartz, and H. E. Stanley, Phys. Rev. A 46, R1724 (1992).
%
\bibitem{Kikkinides:1999ve} E. S. Kikkinides and V. N. Burganos, Phys. Rev. E 59, 7185 (1999).
%
\bibitem{Stanley:1999qf} H. E. Stanley, J. S. Andrade, Jr., S. Havlin, H. A. Makse, and B. Suki, Physica A 266, 5 (1999).
%
\bibitem{Makse:2000bh} H. A. Makse, J. S. Andrade, Jr., and H. E. Stanley, Phys. Rev. E 61, 583 (2000).
%
\bibitem{Araujo:2002dq} A. D. Ara{\'u}jo, A. A. Moreira, H. A. Makse, H. E. Stanley, and J. S. Andrade Jr., Phys. Rev. E 66, 046304 (2002).
%
\bibitem{Araujo:2003cr} A. D. Ara{\'u}jo, A. A. Moreira, R. N. Costa Filho, and J. S. Andrade, Jr., Phys. Rev. E 67, 027102 (2003).
%
\bibitem{Du:1996nx} C. Du, C. Satik, and Y. C. Yortsos, AICHE J. 42, 2392 (1996).
%
\bibitem{Sahimi:1995fk} M. Sahimi, {\it Flow and Transport in Porous Media and
Fractured Rock} (VCH, Weinheim, 1995).
%
\bibitem{Stauffer:1994aa} D. Stauffer and A. Aharony, {\it Introduction to Percolation Theory} (Taylor and Francis, London, 1994).
%
\bibitem{Lee:2009fk} S. B. Lee, Physica A 388, 2271 (2009).
%
\end{thebibliography}
\end{document}